\documentclass[%
showpacs,
 amsmath,amssymb,
 aps,
 twocolumn,
 prl,
 reprint,
floatfix,
]{revtex4-1}

\usepackage{graphicx}
\usepackage{dcolumn}
\usepackage{bm}
\usepackage{color}
\usepackage{acronym}
\usepackage{multirow}
\usepackage{tabularx}
\usepackage{hyperref}
\hypersetup{
  colorlinks=true,        
  linkcolor=black,         
  citecolor=cyan,         
}

\newcommand{\optsnr}{\rho_{\mathrm{opt}}}

\begin{document}

\preprint{APS/123-QED}

%
%
\title{Matching matched filtering with deep networks for gravitational-wave astronomy}

\author{Hunter Gabbard}
 \email{Corresponding author: h.gabbard.1@research.gla.ac.uk}
\author{Michael Williams}
\author{Fergus Hayes}
\author{Chris Messenger}
\affiliation{
 SUPA, School of Physics and Astronomy, \\
 University of Glasgow, \\
 Glasgow G12 8QQ, United Kingdom \\
}

\date{\today}


%
%
\begin{abstract} 
We report on the construction of a deep convolutional neural network that can
reproduce the sensitivity of a matched-filtering search for binary black hole
gravitational-wave signals. The standard method for the detection of well
modeled transient gravitational-wave signals is matched filtering. However, the
computational cost of such searches in low latency will grow dramatically as
the low frequency sensitivity of gravitational-wave detectors improves.
Convolutional neural networks provide a highly computationally efficient method
for signal identification in which the majority of calculations are performed
prior to data taking during a training process. We use only whitened time
series of measured gravitational-wave strain as an input, and we train and test
on simulated binary black hole signals in synthetic Gaussian noise
representative of Advanced LIGO sensitivity. We show that our network can
classify signal from noise with a performance that emulates that of match
filtering applied to the same datasets when considering the sensitivity defined
by Reciever-Operator characteristics.
\end{abstract}



\maketitle

\acrodef{BBH}[BBH]{binary black hole}
\acrodef{SNR}[SNR]{signal-to-noise ratio}
\acrodef{PSD}[PSD]{power spectral density}
\acrodef{FFT}[FFT]{fast Fourier transform}
\acrodef{CNN}[CNN]{convolutional neural network}
\acrodef{ROC}[ROC]{receiver operator characteristic}


%
%

%
%
\textit{Introduction.}--- 
%
%
The field of gravitational-wave astronomy has seen an explosion of compact
binary coalescence detections over the past several years. The first of these
were binary black hole detections~\cite{PhysRevLett.116.061102,
PhysRevLett.116.241103, PhysRevLett.118.221101} and more recently the advanced
detector network made the first detection of a binary neutron star
system~\cite{PhysRevLett.119.161101}. This latter event was seen in conjunction
with a gamma-ray
burst~\cite{2017arXiv171005834L,2017arXiv171005446G,2017arXiv171005449S} and
multiple post-merger electromagnetic signatures~\cite{2017arXiv171005833L}.
These detections were made possible by the Advanced Laser Interferometer
Gravitational wave Observatory (aLIGO) detectors, as well as the recent joint
detections of GW170814 and GW170817 with Advanced Virgo
\cite{PhysRevLett.119.141101,PhysRevLett.119.161101}. Over the coming years
many more such observations, including \ac{BBH}, binary neutron stars, as well
as other more exotic sources are likely to be observed on a more frequent basis. As
such, the need for more efficient search methods will be more pertinent as the
detectors increase in sensitivity.

%
%
The algorithms used by the search pipelines to make
detections~\cite{0264-9381-33-21-215004, 0004-637X-748-2-136,
PhysRevD.90.082004} are, in general, computationally expensive. The methods
used are complex, sophisticated processes computed over a large parameter space
using advanced signal processing techniques. The computational cost to run the
search analysis is due to the large parameter space and the increasing cost of
analysing longer duration waveforms as the low frequency sensitivity of the
detectors improves. Distinguishing noise from signal in these search pipelines
is acheived using a technique known as template based matched-filtering. 

%
%
Matched-filtering uses a bank~\cite{PhysRevD.86.084017,
1705.01845,PhysRevD.80.104014, PhysRevD.90.082004, PhysRevD.89.084041} of
template waveforms~\cite{PhysRevD.44.3819, PhysRevD.89.061502,
PhysRevD.89.024003, Blanchet2014} each with different component mass components
and/or spin values. A template bank will span a large astrophysical parameter
space since we do not know a priori the true gravitational-waves parameter
values. Waveform models that cover the inspiral, merger, and ringdown phases of
a compact binary coalescence are based on combining post-Newtonian
theory~\cite{PhysRevD.84.049901,PhysRevD.80.084043,Blanchet2014,PhysRevD.93.084054},
the effective-one-body formalism~\cite{PhysRevD.59.084006}, and numerical
relativity simulations~\cite{PhysRevLett.95.121101}.

%
%
Deep learning is a subset of machine learning which has gained in popularity in
recent years~\cite{NIPS2012_4824, 1406.2661, 1409.1556, 1412.7062, 1311.2901,
1409.4842} with the rapid development of graphics-processing-unit technology.
Some successful implementations of deep learning include image processing~\cite{1603.08511,1412.2306,NIPS2012_4824}, medical
diagnosis~\cite{KONONENKO200189}, and microarray gene expression
classification~\cite{Pirooznia2008}. There has also been
some recent success in the field of gravitational-wave astronomy in the form of
glitch classification \cite{PhysRevD.95.104059,
0264-9381-34-6-064003,1706.07446} and notably for signal
identification~\cite{1701.00008} where it was first shown that deep learning
could be a detection tool. Deep learning is able to
perform analyses rapidly since the method's computationally intensive
stage is pre-computed during the training prior to the analysis of actual data.
This results in low-latency searches that can be orders of magnitude
faster than other comparable classification methods. 

%
%
A deep learning algorithm is composed of stacked arrays of processing units,
called neurons, which can be from one to several layers deep. A neuron
acts as a filter, whereby it performs a transformation on an array of inputs.
This transformation is a linear operation between the input array and the
weight and bias parameters assosicated to the neuron. The resulting array is
then typically passed to a non-linear activation function to constrain the
neuron output to be within a set range. Deep learning algorithms typically
consist of an input layer, followed by one to several hidden layers and then
one to multiple output neurons. The scalars produced from the output neurons
can be used to solve classification problems, where each output neuron
corresponds to the probability that an input sample is of a certain
class.

%
%
In this letter we investigate the simplest case of establishing whether a
signal is present in the data or if the data contains only detector noise. We
propose a deep learning procedure requiring only the raw data time series as
input with minimal signal pre-processing. We compare the results of our network
with the widely used matched-filtering technique and show how a deep learning
approach can be pre-trained using simulated data-sets and applied in
low-latency to achieve the same sensitivity as established matched-filtering
techniques. 

%
%
\textit{Simulation details.}--- 
%
%
In order to make a clean comparison between deep learning approach and
matched-filtering, we distinguish between two cases, \ac{BBH} merger
signals in additive Gaussian noise (signal+noise) and Gaussian noise alone
(noise-only). We choose to focus on \ac{BBH} signals rather than including
binary neutron star systems for the reason that \ac{BBH} systems are higher
mass systems and have shorter duration signals once the inspiralling systems
have entered the Advanced LIGO frequency band. They typically then merge on the
timescale of $\mathcal{O}(1)$ sec allowing us to use relatively small datasets
for this study. 

%
%
The input datasets consist of ``whitened'' simulated gravitational-wave
timeseries where the whitening process uses the detector noise \ac{PSD} to  
rescale the noise contribution at each frequency to have equal power. 
Our noise is initially generated from a \ac{PSD} equivelent to the Advanced
LIGO design sensitivity~\cite{2016LRR....19....1A}. 

%
%
Signals are simulated using a library of gravitational-wave data analysis
routines called \texttt{LALSuite}. We use the IMRPhenomD type
waveform~\cite{PhysRevD.93.044006, PhysRevD.93.044007} which models the
inspiral, merger and ringdown components of \ac{BBH} gravitational-wave
signals. We simulate systems with component black hole masses in the range from
5\(M_\odot\) to 95\(M_\odot\), $m_{1} > m_{2}$, with zero spin.
Training, validation and testing datasets contain signals drawn from an
astrophysically motivated distribution where we assume
$m_{1,2}\sim\log{m_{1,2}}$~\cite{PhysRevX.6.041015}. Each signal is given a
random right ascension and declination assuming an isotropic prior on the sky,
the polarization angle and phase are drawn from a uniform prior on the range
$[0,2\pi]$, and the inclination angle is drawn such that the cosine of
inclination is uniform on the range $[-1,1]$. The waveforms are then randomly
placed within the time series such that the peak amplitude of each waveform is
randomly positioned within the fractional range $[0.75,0.95]$ of the
timeseries. 

\begin{figure} 
\includegraphics[width=\columnwidth]{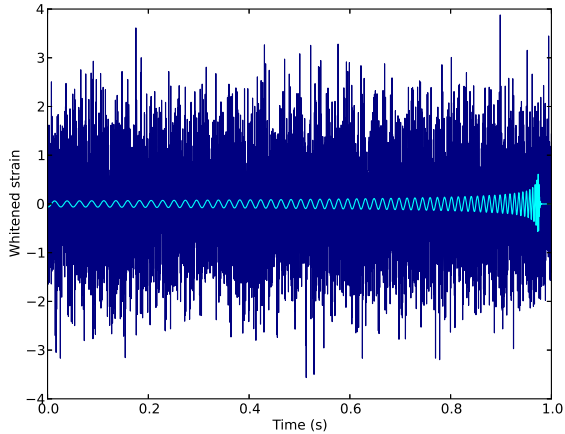}
\caption{A whitened noise-free timeseries of a \ac{BBH} signal with component
masses $m_{1}=12.06\mathrm{M}_{\odot}$ and $m_{2}=7.54\mathrm{M}_{\odot}$ with
optimal \ac{SNR} $\optsnr=8$ (cyan). The dark blue timeseries shows the same
gravitational-wave signal with additive whitened Gaussian noise of unit
variance. This latter timeseries is representative of the datasets used to
train, validate, and test the deep neural network.\label{fig:waveform}}
\end{figure}

%
%
The waveform amplitude is scaled to achieve a predefined optimal \ac{SNR}
defined as
%
%
\begin{equation}\label{eq:snr} 
\rho_{\mathrm{opt}}^{2} = 4
\int\limits_{f_{\mathrm{min}}}^{\infty} \frac{\lvert
\tilde{h(f)}\rvert^{2}}{S_{\mathrm{n}}(f)} df,
\end{equation}
where $\tilde{h(f)}$ is the frequency domain representation of the
gravitational-wave strain and $S_{\mathrm{n}}(f)$ is the detector noise
\ac{PSD}. The simulated time series were chosen to be 1~sec in duration sampled
at 8192 Hz. Therefore we consider $f_{\mathrm{min}}$ as the frequency of the
gravitational-wave signal at the start of the sample timeseries. An example
timeseries can be seen in Fig.~\ref{fig:waveform}. 

%
%
Due to the requirements of the matched-filtering comparison it was necessary to
add padding to each timeseries so as to avoid edge effects. Therefore
each 1~sec timeseries has an additional 0.5~sec of data prior to and after the signal.
The signal itself has a Tukey window ($\alpha=1/8$) applied to truncate the
signal content to the central 1~sec. The \ac{CNN} approach only has access to
this central 1~sec of data. Similarly, the optimal \ac{SNR} is computed considering
only the central 1~sec.

%
%
Supervised deep learning requires datasets to be sub-divided into training,
validation, and testing sets. Training sets are the data samples that the
network learns from, the validation set allows the developer to verify that the
network is learning correctly, and the test set is used to quantify the
performance of the trained network.  Of the dataset generated we use $90\%$ of
these samples for training, $5\%$ for validation, and $5\%$ for testing. A
dataset was generated for each predefined optimal \ac{SNR} value ranging
from $1$--$10$ in integer steps. 

%
%
Our training datasets contain $4\times 10^{5}$ independent timeseries with 50\%
containing signal+noise and 50\% noise-only. For each simulated
gravitational-wave signal (drawn from the signal parameter space) we generate
25 independent noise realizations from which 25 signal+noise samples are
produced. This procedure is standard within machine learning classification and
allows the network to learn how to identify individual signals under different
noise scenarios. Each noise-only sample consists of an independent noise
realization and in total we therefore use 1000 unique waveforms in the
$m_{1},m_{2}$ mass space. Each data sample timeseries is then represented in
the form of a $1 \times 8192$ pixel image with the gray-scale intensity of each
pixel proportional to the gravitational-wave amplitude.

%
%

%
%
\textit{The Deep Network approach.}--- 
%
%
In our model, we use a variant of a deep learning algorithm called a
\ac{CNN}~\cite{726791} composed of multiple layers. The input layer holds 
the raw pixel values of the sample image which, in
our case, is a 1-dimensional timeseries vector. The weight and bias parameters 
of the network are also in 1-dimensional vector
form. Each neuron in the convolutional layer computes the convolution between
the neuron's weight vector and the outputs from the layer below it, and then
the result is summed with the bias vector. Neuron weight vectors are updated
through an optimisation algorithm called back-propogation~\cite{LeCun1998}.
Activation functions apply an element-wise non-linear operation rescaling their
inputs onto a specific range and leaving the size of the previous layer's
output unchanged. Pooling layers perform a downsampling operation along the
spatial dimensions of their input. Finally we have a hidden layer connected to an output layer which
computes the inferred class probabilities. These values are input to a loss
function, chosen as the binary cross-entropy~\cite{tensorflow2015-whitepaper},
defined as
%
%
\begin{equation} \label{eq:loss} 
f(\theta) =
-\sum_{i\in\text{S}}\mathrm{log}(\theta^{\text{S}}_{i})-\sum_{i\in\text{N}}\mathrm{log}(\theta^{\text{N}}_{i}), 
\end{equation}
where $\theta^{\text{S/N}}_{i}$ is the predicted probability of class
signal+noise (S) or noise-only (N) for the $i$'th training sample. The
loss function is minimised when input data samples are assigned the correct
class with the highest confidence. 

%
%
In order to optimise a network, multiple hyper-parameters must be tuned.  We
define hyper-parameters as parameters we are free to choose. Such
parameters include the number and type of network layers, the
number of neurons within each layer, size of the neuron weight vectors, 
max-pooling parameters, type of activation functions, preprocessing of
input data, learning rate, and the application (or otherwise) of specific deep
learning techniques. We begin the process with the simplest network that
provides a discernible level of effective classification. In most cases this
consists of an input, convolutional, hidden, and logistic output layer.
The optimal network structure was determined through multiple
tests and tunings of hyperparameters by means of trial and error.

%
%

%
%
During the training stage an optimization function (back-propagation) works by
computing the gradient of the loss function (Eq.~\ref{eq:loss}), then
attempting to minimize that loss function. The errors are then propagated back
through the network while also updating the weight and bias terms
accordingly.  Back propagation is done over multiple iterations called epochs.
We use adaptive moment estimation with incorporated Nesterov
momentum~\cite{dozat2016incorporating} with a learning rate of $0.002$,
$\beta_{1}=0.9$, $\beta_{2}=0.999$, $\epsilon = 10^{-8}$ and a momentum
schedule of $0.004$. We outline the structure of the final neural network
architecture in Table~\ref{table:network}.



%
%
The final ranking statistic that we extract from the \ac{CNN} analysis is taken
from the output layer, composed of 2 neurons, where each neuron will
produce a probability value between 0 and 1 with their sum being unity. Each
neuron gives the inferred probability that the input data belongs to the noise
or signal+noise class respectively.

%
%

\begin{table}[]
\begin{tabular}{lccccccccc}
\hline
\hline
Parameter & \multicolumn{9}{c}{Layer}\\
\cline{2-10}
(Option) & 1 & 2 & 3 & 4 & 5 & 6 & 7 & 8 & 9 \\
\hline
Type & C & C & C & C & C & C & H & H & H \\
No. Neurons  & 8  & 8  & 16 & 16 & 32 & 32 & 64  & 64  & 2  \\
Filter Size  & 64 & 32 & 32 & 16 & 16  & 16  & n/a & n/a & n/a  \\
MaxPool Size & n/a & 8 & n/a & 6 & n/a & 4 & n/a & n/a & n/a \\
Drop out  & 0 & 0 & 0 & 0 & 0 & 0 & 0.5 & 0.5 & 0 \\
Act. Func. & Elu & Elu & Elu & Elu & Elu & Elu & Elu & Elu & SMax \\
\hline
\end{tabular}
\caption{The optimised network consisting of 6 convolutional layers (C),
followed by 3 hidden layers (H). Max-pooling is performed on the first, fifth,
and eighth layer, whereas dropout is only performed on the two hidden layers.
Each layer uses an exponential linear unit (Elu) activation function (with
range $[-1,\infty]$) while the last layer uses a Softmax (SMax) activation
function in order to normalize the output values to be between zero and one so
as to give a probability value for each class.\label{table:network}}
\end{table}

%
%
\textit{Applying matched-filtering.}---
%
%
In order to establish the power of the deep learning approach we must compare
our results to the standard matched-filtering process used in the detection of
compact binary coalescence
signals~\cite{PhysRevD.85.122006,2013PhRvD..87b4033B}. The ranking statistic
used in this case is the matched-filter \ac{SNR} numerically maximized over
arrival time, phase and distance. By first defining the noise weighted inner
product as a function of a time shift $\Delta t$ between the arrival time of
the signal and the template,
%
%
\begin{equation}\label{eq:inner}
(a\mid b)[\Delta t] =
4\int_{f_{\mathrm{min}}}^{\infty}\frac{\tilde{a}(f)\tilde{b}^{*}(f)}{S_{\mathrm{n}}(f)}e^{2\pi i
f\Delta t}\,df,
\end{equation}
we can construct the matched-filter \ac{SNR} as 
\begin{equation}
\rho^{2}[\Delta t]=\frac{(s\mid h)^{2}[\Delta t] + i(s\mid h)^{2}[\Delta t]}{(h\mid h)}
\end{equation}
where $s$ is the data containing noise and a potential signal, and $h$ is the
noise-free gravitational-wave template. For a given template this quantity is
efficiently computed using the \ac{FFT} and the \ac{SNR} timeseries maximised
over $\Delta t$. The subsequent step is to further numerically maximize this
quantity over a collection of component mass combinations. In this analysis a
comprehensive template bank is generated in the $m_{1},m_{2}$ mass space
covering our predefined range of masses. We use a maximum mismatch of $3\%$ and
a lower frequency cutoff of $20~\mathrm{Hz}$ using the PyCBC geometric
non-spinning template bank generation
tool~\cite{pycbc-software,0264-9381-33-21-215004}. This template bank contained
$8056$ individual templates. 

%
%
When generating an \ac{SNR} timeseries for an input dataset we select
$f_{\mathrm{min}}$ according to the conservative case (lowest
$f_{\mathrm{min}}$) in which the signal merger occurs at the 0.95 fraction of
1~sec timeseries. We therefore select only maximised \ac{SNR} timeseries
values recovered from within the $[0.75,0.95]$ fractional range since this is
the parameter space on which the \ac{CNN} has been trained. For the practical
computation of the matched-filtering analysis we take each of the data samples
from the testing dataset to compute the matched-filter ranking statistic.

%
%
%
%

%
%
\textit{Results.}---
%
%
After tuning the multiple hyper-parameters (Table~\ref{table:network}) and
training the neural network, we present the results of our \ac{CNN} classifier
on a noise versus signal+noise sample set. With values of statistics now
assigned to each test data sample from both the \ac{CNN} and matched-filtering
approaches, and having knowledge of the true class associated with each sample,
we may now construct \ac{ROC} curves. 

%
%

\begin{figure}[]
\includegraphics[width=\columnwidth] {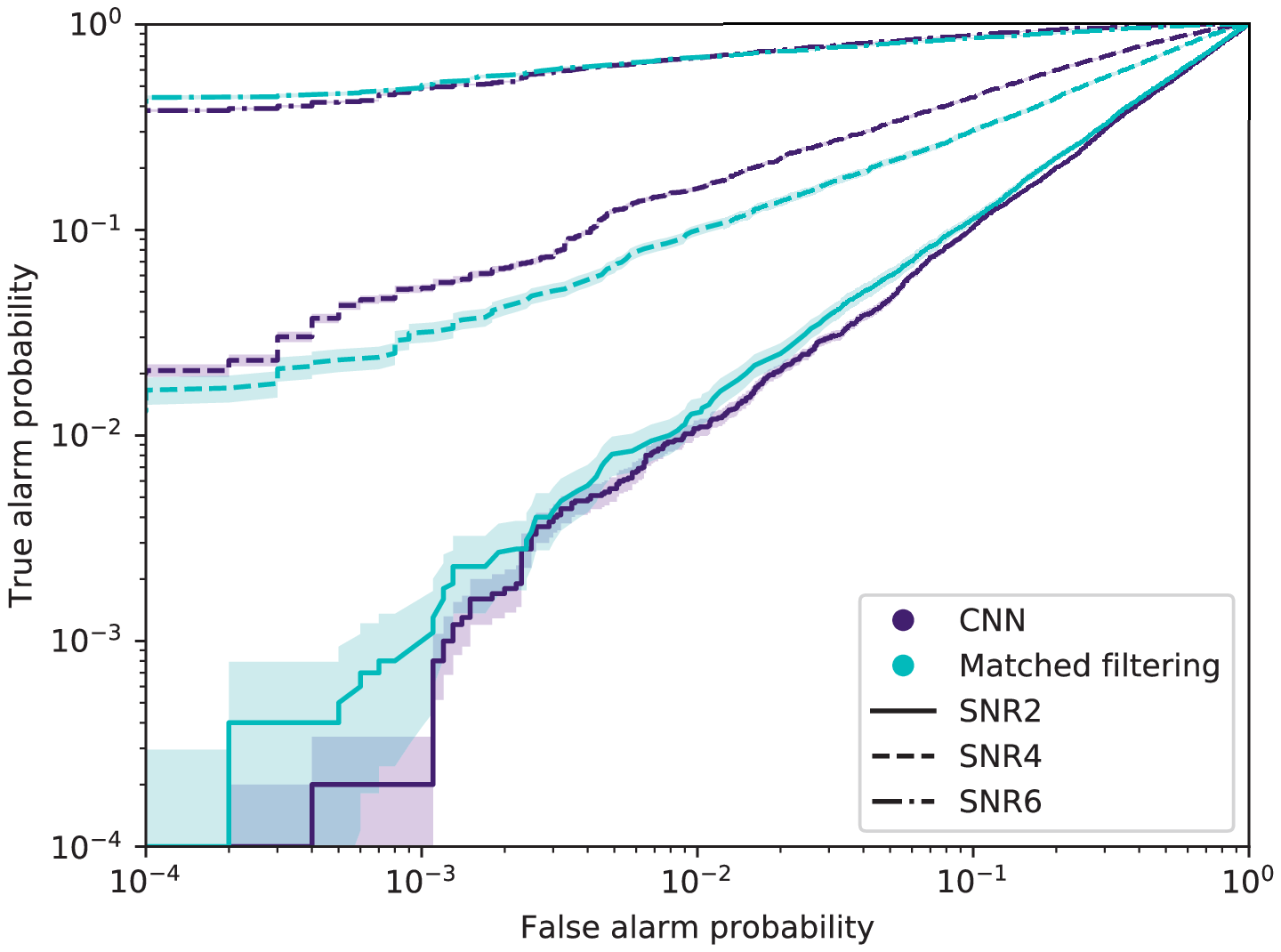}
\caption{The \ac{ROC} curves for test datasets containing signals with optimal
\ac{SNR}, $\rho_{\mathrm{opt}}=2,4,6$. We plot the true alarm probability
versus the false alarm probability estimated from the output of the \ac{CNN}
(purple) and matched-filtering (cyan) approaches. Uncertainties in the true
alarm probability correspond to 1-$\sigma$ bounds assuming a binomial
distribution.} \label{fig:ROC_curves} 
\end{figure}

%
%
In Fig.~\ref{fig:ROC_curves} we compare our \ac{CNN} results to that of
matched-filtering. Given the ranking statistic from a particular analysis and
defining a parametric threshold value on that statistic we are able to plot the
fraction of noise samples incorrectly identified as signals (false alarm
probability) versus the fraction of signal samples correctly identified (true
alarm probability). These curves are defined as \ac{ROC} curves and a ranking
statistic is deemed superior to another if at a given false alarm probability
it achieves a higher detection probability. Our results show that the \ac{CNN}
approach closely matches the sensitivity of matched-filtering for all test
datasets across the range of false alarm probabilities explored in this
analysis\footnote{We are limited to a minimal false alarm probability of $\sim
10^{-4}$ due to the limited number of testing samples used.}.

%
%

\begin{figure}[]
\includegraphics[width=\columnwidth] {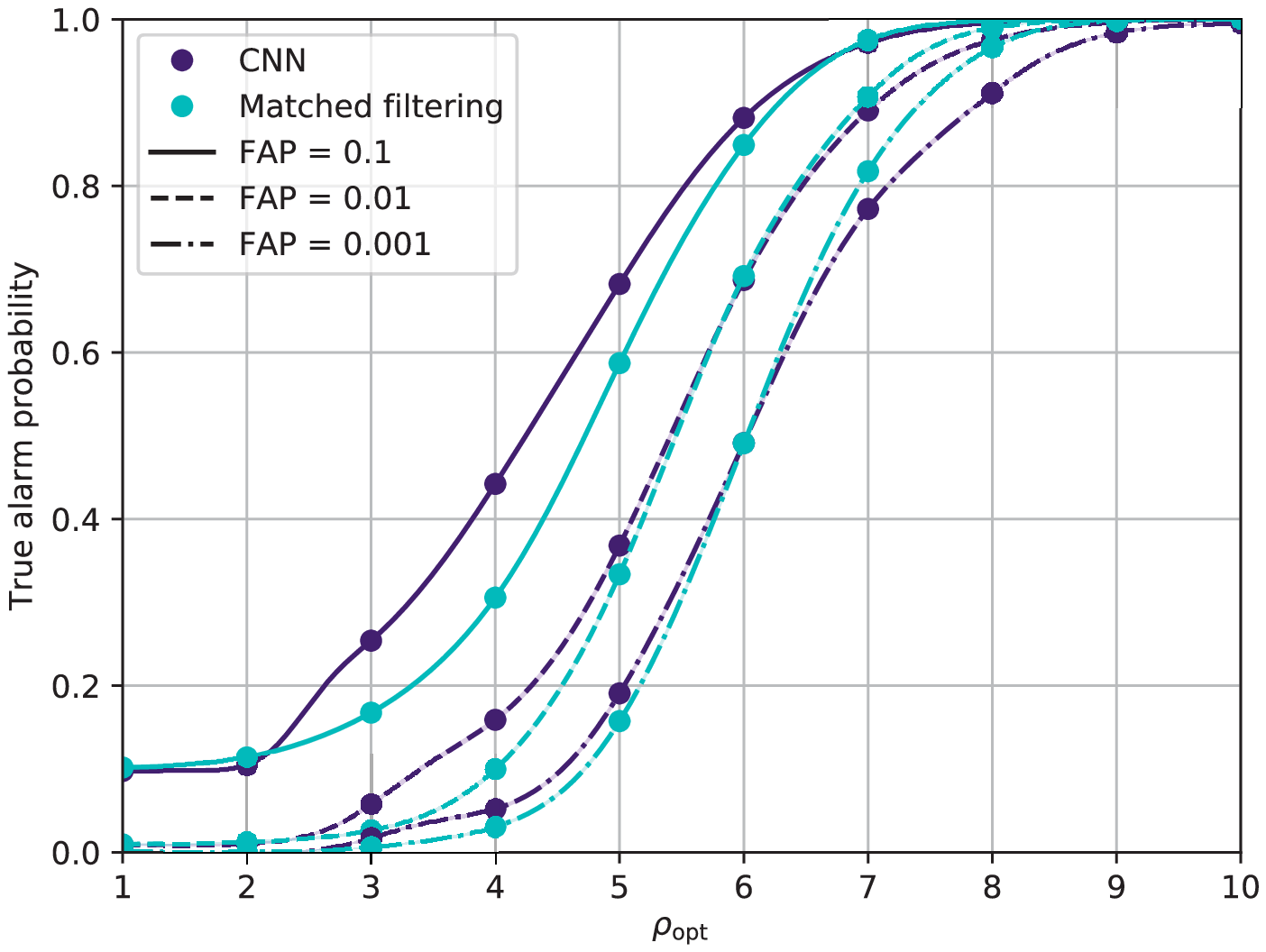}
\caption{Efficiency curves comparing the performance of the \ac{CNN} and
matched-filtering approaches for false alarm probabilities $10^{-1}$ (solid),
$10^{-2}$ (dashed), $10^{-3}$ (dot-dashed). The true alarm probability is
plotted as a function of the optimal \ac{SNR} for the \ac{CNN} (purple) and the
matched-filtering (cyan) analyses. Solid dots indicate at which \ac{SNR} values
analyses were performed and line thicknesses are indicative of the
statistical uncertainties in the curves.\label{fig:efficiency_curve}} 
\end{figure}

%
%
We can make an additional direct comparison between approaches by fixing a
false alarm probability and plotting the corresponding true alarm probability
versus the optimal \ac{SNR} of the signals in each test dataset. We show these
efficiency curves in Fig.~\ref{fig:efficiency_curve} at false alarm
probabilities $10^{-1},10^{-2},10^{-3}$ for both the \ac{CNN} and
matched-filtering approaches. We again see very good agreement between the
approaches at all false alarm probabilities with the \ac{CNN} sensitivity
exceeding that of the matched-filter approach at low \ac{SNR} and high false
alarm probability. Conversely we see the matched-filter sensitivity marginally
exceeds the \ac{CNN} at high \ac{SNR} and low false alarm probability. This
latter discrepancy can be mitigated by increasing the number of training
samples.


%
%

%
%
\textit{Conclusions.}--- 
%
%
We have demonstrated that deep learning, when applied to
gravitational-wave timeseries data, is able to closely reproduce the
results of a matched-filtering analysis in Gaussian noise. We employ a deep
convolutional neural network with rigorously tuned hyperparameters and produce
an output that returns a ranking statistic equivalent to the inferred
probability that data contains a signal. Matched-filtering analyses are often
described as the optimal approach for signal detection in Gaussian noise. By
building a neural network that is capable of reproducing this
optimality we answer a fundamental question regarding the applicability of
neural networks for gravitational-wave data analysis. 

%
%
In practice, searches for transient signals in gravitational-wave data are
strongly affected by non-Gaussian noise artefacts. To account for this,
standard matched-filtering approaches are modified to include carefully chosen
changes to the ranking statistic~\cite{PhysRevD.71.062001,0004-637X-849-2-118}
together with the excision of poor quality data~\cite{1710.02185,
0264-9381-33-13-134001}. Our analysis represents a starting point from which a
deep network can be trained on realistic non-Gaussian data. Since the claim of
matched-filtering optimality is applicable only in the Gaussian noise case,
there exists the potential for deep networks to exceed the sensitivity of
existing matched-filtering approaches in real data.

%
%
In this work we have presented results for \ac{BBH} mergers, however, this
method could be applied to other merger types, such as binary neutron star and
neutron star-black hole signals. This supervised learning approach can also be
extended to other well modelled gravitational-wave targets such as the
continuous emission from rapidly rotating non-axisymmetric neutron
stars~\cite{1707.02669}. 
Finally we mention the possibilities for parameter
estimation~\cite{1701.00008} where in the simplest cases an output regression
layer can return point estimates of parameter values. As was exemplified in the
case of GW170817, rapid detection confidence coupled with robust and equally
rapid parameter estimates is critical for gravitational-wave multi-messenger
astronomy. 

%
%
\emph{Acknowledgements.}---
We would like to acknowledge valuable input from the LIGO-Virgo Collaboration
specifically from T.~Dent, M.~Cavalgia, and the compact binary coalescence and
machine-learning working groups. The authors also gratefully acknowledge the
Science and Technology Facilities Council of the United Kingdom. CM is
supported by the Science and Technology Research Council (grant
No.~ST/~L000946/1).
%


\bibliographystyle{apsrev4-1}
\bibliography{references}

\end{document}